\documentclass[11pt, a4paper]{article} 
\usepackage{latexsym}
\usepackage{epsfig}
\usepackage{amssymb}
\usepackage{amsthm}
\usepackage{amsmath}
\usepackage{amssymb,amsfonts}
\usepackage{verbatim}

\newcommand{\be}{\begin{eqnarray}}
\newcommand{\ee}{\end{eqnarray}}

\usepackage{jheppub}
\usepackage{epsfig,multicol,bbm}

\newcommand\fverb{\setbox\fverbbox=\hbox\bgroup\verb}
\newcommand\fverbdo{\egroup\medskip\noindent%
            \fbox{\unhbox\fverbbox}\ }
\newcommand\fverbit{\egroup\item[\fbox{\unhbox\fverbbox}]}
\newbox\fverbbox

\subheader{LMU-ASC 16/11}
\title{UV-protected (Natural) Inflation:\\ Primordial Fluctuations and non-Gaussian Features}

\author{Cristiano Germani}
\author{and Yuki Watanabe}

\affiliation{Arnold Sommerfeld Center for Theoretical Physics, Ludwig-Maximilians-University, Theresienstrasse 37, 80333 Munich, Germany}

\emailAdd{cristiano.germani@physik.lmu.de}
\emailAdd{yuki.watanabe@physik.lmu.de}

\abstract{We consider the UV-protected inflation, where the inflaton potential is obtained by quantum (one-loop) breaking of a global symmetry into a discrete symmetry. In this model, all coupling scales are sub-Planckian. This is achieved by coupling the inflaton kinetic term to the Einstein tensor such that the friction is enhanced gravitationally at high energies. In this respect, this new interaction makes virtually any potential adequate for inflation while keeping the system perturbative unitary. We show that even if the gravitationally enhanced friction intrinsically contains new nonlinearities, the UV-protected inflation (and any similar models) behaves as a single field scenario with red tilted spectrum and potentially detectable gravitational waves. Interestingly enough, we find that non-Gaussianity of the curvature perturbations in the local form are completely dominated by the nonlinear gauge transformation from the spatially flat to uniform-field gauge and/or by parity violating interactions of the inflaton and gauge bosons. In particular, the parity violating interactions may produce detectable non-Gaussianity. }


\begin{document}

\maketitle  

\section{Introduction}

The observed fluctuations of temperature of the Universe \cite{Komatsu:2010fb}, the Cosmic Microwave Background Radiation (CMBR), is spectacularly explained by postulating inflation. In its simplest form, inflation is an early time exponential expansion of the Universe driven by a scalar field $\phi$ slowly rolling in its own potential $V(\phi)$. 
Quantum fluctuations of the inflaton field $\phi$ produce little glitches of inhomogeneity that are then mapped into the CMBR. One of the biggest (generic) successes of inflation is to predict that the statistical distribution of the CMBR is almost Gaussian, because of its quantum origin \cite{Mukhanov:1981xt}.

Canonical inflation, however, {\it may} suffer from the ``eta'' problem: 
In order to obtain an exponential (DeSitter) expansion of the early Universe, the inflaton field value must generically run over trans-Planckian scales. In this case, one should consider at least the following corrections to the potential
\be
V=V_0\left[1+\sum_{k=1}^{\infty} c_k\left(\frac{\lambda\phi}{M_p}\right)^k\right]\ ,
\ee
where $V_0$ is the bare potential, $c_k$ are constants encoding the quantum gravity corrections and $\lambda\phi$ is some physical quantity constructed from $\phi$. Here $\lambda$ is not necessarily constant but may generically depend on $\phi$ itself or other characteristic scales of the inflating background. Note indeed that the bare $\phi$ has no physical meaning as it is {\it not} any measurable quantity \cite{Linde:1990}. Unless $\lambda\phi\ll M_p$ (even for $\phi\gg M_p$),\footnote{We thank Andrei Linde for pointing this out, see Sec. 2.4 of \cite{Linde:1990}.} or unless new symmetries are introduced, one would need infinite fine tuning to get $V\sim V_0$ so to trust the bare potential during inflation. This is the feared ``eta'' problem.

In this paper, we will consider the recently proposed ``UV-Protected'' inflationary scenario of \cite{Germani:2010hd} where, thanks to the ``gravitationally enhanced friction'' (GEF) mechanism firstly introduced in the ``New Higgs'' Inflation \cite{Germani:2010gm}, all scales of the Natural Inflation Lagrangian \cite{Freese:1990rb} can be consistently taken to be sub-Planckian, without spoiling the inflationary trajectory.
In the UV-protected inflation, the inflaton potential is a one-loop effect coming from quantum breaking of a global (shift) symmetry into a discrete symmetry, thus, all quantum corrections on the potential are already incorporated and under-control.

The GEF is a mechanism that drastically increases the (gravitational) friction acting on a scalar field. This mechanism makes even too steep potentials that would not inflate enough in General Relativity (GR) otherwise, adequate for inflation. An important property of this mechanism is that it is realized without introducing any degree of freedom more than the inflaton and the massless graviton. 

As we shall see, the GEF mechanism is realized by a nonminimal coupling of the kinetic term of the scalar field with the Einstein tensor $G_{\alpha\beta}$ of the form\footnote{Horndeski \cite{Horndeski:1974} has first considered the most general nonminimal interactions of the scalar field and gravity, including this interaction, to give second order field equations in four dimensions. This theory has recently been rediscovered by \cite{Deffayet:2011gz} in arbitrary dimensions (see Appendix  A of \cite{Kobayashi:2011nu} for equivalence of the theories). Amendola \cite{Amendola:1993uh} was the first considering nonminimally derivative coupled scalar field theories to gravity in the context of cosmology.} 
\be\label{int}
{\cal L}=-\frac{1}{2}{\sqrt{-g} }\left(g^{\alpha\beta}-\frac{G^{\alpha\beta}}{M^2}\right)\partial_\alpha\phi\partial_\beta\phi\ .
\ee
Generically, one would expect that the introduction of a new mass scale, $M$, and a new nonlinear interaction of gravity to the scalar field produces non-Gaussian fluctuations larger than those in GR (given the same potential $V(\phi)$) during inflation. Although we do expect new features in the gravitational wave sector, we will show that non-Gaussianities are actually still suppressed in the scalar sector. 

This result has a very interesting physical explanation: during inflation, $G^{\alpha\beta}/M^2\simeq -3 (H^2/M^2) g^{\alpha\beta}$ and therefore, by canonical normalization of the inflaton (at lowest order in slow roll), the scale $M$ gets absorbed into redefinitions of the slow roll parameters. Thus, the model resembles a single scalar field scenario with no extra scale. In this case, non-Gaussianities of curvature perturbations are suppressed by slow-roll as in \cite{Maldacena:2002vr}.

We work with the metric signature $(-,+,+,+)$, $R^{\alpha}{}_{\beta\gamma\delta}=\Gamma^{\alpha}{}_{\beta\delta,\gamma}-\Gamma^{\alpha}{}_{\beta\gamma,\delta}+\Gamma^{\alpha}{}_{\lambda\gamma}\Gamma^{\lambda}{}_{\beta\delta}- \Gamma^{\alpha}{}_{\lambda\delta}\Gamma^{\lambda}{}_{\beta\gamma}$ and $R_{\mu\nu}=R^{\alpha}{}_{\mu\alpha\nu}$, which agree with \cite{Misner:1974qy, Wald:1984rg, Liddle:2000cg}.
We assume the Einstein summation convention over the spacetime and space indices, where Greek letters, $\alpha,\ \beta,\ \cdots$, run over the spacetime indices while Latin letters, $i, \ j,\ \cdots$, run over the space indices. We employ natural units $\hbar = c = 1$ and the reduced Planck mass scale $M_p = (8\pi G)^{-1/2} = 2.4 \times 10^{18}$ GeV. 

\section{Gravitationally enhanced friction (GEF) in a nutshell}

Inflation is a simple mechanism for providing an exponential (DeSitter)
expansion of the Early Universe. In there, gravity is sourced by a slow rolling scalar field $\phi$ with potential $V$, namely $\dot{\phi}^2\ll V$.

An obvious fulfillment of slow roll is that $V$ has a non-trivial 
(positive) minimum.  However, in this case the exponential expansion of the Universe will
 never end and structures (like galaxies etc.) will never form. What we then would like to achieve is a slow 
motion of the scalar field, far from the equilibrium points of the potential $V$. 

Let us ignore gravity for a moment and gain some intuition by considering a scalar field in one dimension. Our optimal goal would be that the scalar field energy ($E$) is dominated by the potential energy, i.e. we want 
\be\label{e}
E\simeq V\ ,
\ee
in a non-equilibrium point of $V$. A typical way for this to be true is that the scalar field experiences large friction while rolling down the potential. 
However, in this case the energy of the isolated scalar field is {\it not} a conserved quantity ($\dot E\neq 0$). Nevertheless, we will look for a solution that almost has a constant energy so that (\ref{e}) will be satisfied not only instantaneously but also for a long period. In other words, we want (we use natural units)
 \be\label{ep}
 \tilde\epsilon\equiv-\frac{\dot E}{E^2}\ll 1\ ,
 \ee
 for a long time, i.e. we need the second condition 
 \be\label{et}
 \tilde\delta\equiv\frac{\dot{\tilde\epsilon}}{\tilde\epsilon E}\ll 1,
 \ee
 to be fulfilled.
 
In order to be effective, the friction term ($\propto \dot\phi$) must dominate over the acceleration. Let us parametrize it by $\tilde\mu\dot\phi$. The force pulling the scalar field down the potential is the gradient of the potential with respect to the field. We then have that, if friction dominates, the field equation is
\be\label{friction}
\tilde\mu\dot\phi\simeq-V'\ ,
\ee
where $f'=df/d\phi$.

In order to have an almost constant energy, the friction coefficient must also be roughly constant. In this case, the parameters (\ref{ep}) and (\ref{et}) will look like
\be\label{cond}
\tilde\epsilon\simeq \frac{V'^2}{V^2}\frac{1}{\tilde \mu}\ ,~~~~~\tilde\delta\simeq-2\frac{V''}{V}\frac{1}{\tilde \mu}+2 \tilde\epsilon\ .
\ee
We see that both $\tilde\epsilon,\tilde\delta \propto \tilde\mu^{-1}$; as anticipated, in absence of friction ($\tilde\mu\rightarrow 0$) slow roll is impossible unless $V'\sim V''\sim 0$.

As we noticed, in order to produce a slow roll, the friction must be roughly time independent in this regime. 
We have two possibilities. The trivial one is that $\tilde\mu={\rm const}$ and the more interesting one is that $\tilde\mu=\tilde\mu(E)$. Only the second one fulfills the requirement to have an inefficient friction at low energies, or in the language of inflation, a graceful exit from inflation (remember we want galaxies at the end!). Moreover, if we want $\tilde\mu$ to dominate at high energies, we need it to be a growing function of $E$. 

There are two ways that the ``friction'' may appear in the full equation of motion of the scalar field. The first one is 
\be\label{a}
\ddot\phi+\tilde\mu\dot\phi=-V'\ .
\ee
This case is well known and it implies that the scalar field dissipates on other fields ($\tilde\mu$ is here the decay rate). This means that in order to implement (\ref{a}) we would need to introduce new degrees of freedom. Another way to see that is to check the condition of the friction domination over the acceleration, i.e.
\be
\tilde{\tilde\delta}\equiv\frac{\ddot\phi}{\tilde\mu\dot\phi}\ll 1\ .
\ee
One can show that $\tilde{\tilde\delta}\propto\tilde\delta (E/\tilde\mu)$, and therefore a new parameter controlling the system must be introduced if $\tilde\mu\not\propto E$. This automatically implies, as said, the presence of new degrees of freedom.

Another choice to implement the friction would be the following: Let us define for dimensional reasons, $\tilde\mu=3E\mu$. We may consider the following equation
\be\label{b}
\mu\left(\ddot\phi+3E\dot\phi\right)=-V'\ .
\ee
We find that, in this case, $\tilde{\tilde\delta}\sim \tilde{\delta}$. This is a good starting point to avoid the introduction of new degrees of freedom as no new parameters are needed to control (\ref{b}). 

Let us now introduce gravity and the physical four dimensions. In this case one should consider energy density instead of energy. Although the previous discussion changes in details, it does not alter the physical intuition. The gravitational Hamiltonian density (${\cal H}$) in the Friedmann-Robertson-Walker (FRW) Universe is $3M_p^2H^2$ where $H=\dot a/a$ \cite{Wald:1984rg}. We may then consider ${\cal H}$ in the left hand side of (\ref{e}) and $H$ instead of $E$ anywhere else. Note that the conditions (\ref{cond}) also change, however, the discussion about the degree of freedom remains  unchanged. The incarnation of the condition (\ref{e}) is therefore the Friedmann equation $3M_p^2H^2\simeq V$ and $\tilde\mu=3H\mu(H)$. 

Note that, in the limit in which $\dot\mu(H)\simeq 0$, $\mu$ can be absorbed in a time rescaling, i.e., the scalar field experiences, during slow-roll, an effective time $t_{\rm eff}=t/\sqrt{\mu}$. In the limit in which $\mu\rightarrow\infty$ the scalar field gets frozen. In other words, the physical intuition of the case (\ref{b}) is that scalar field's clock is moving slower than observer's clock!

The expansion of the Universe acts as a friction term for the scalar field due to the redshift effect. 
In this case, the friction is just the simplest case of (\ref{b}) where $\mu=1$. Can we be more sophisticated than that without introducing new degree of freedom and in a covariant way? The answer is yes \cite{Germani:2010gm,Germani:2010hd}. 

First of all, one may ask why to bother. Suppose that we have a potential $V$ too steep to produce inflation with just the help of gravitational friction ($\mu=1$). 
For example, this is the case in which the Higgs boson is considered as an inflaton. Then, if $\mu\gg 1$, the same (steep) potential $V$ can be turned to be adequate for inflation.

As we discussed, we need $\tilde\mu$ to grow with energy and be positive. The simplest case is then 
\be\label{mu}
\mu=\left(1+\frac{3H^2}{M^2}\right)\ ,
\ee
 where $M$ is some energy scale. Now if during inflation $H^2\gg M^2$, our goal to enhance friction is achieved and slow roll is easier to obtain.

The questions are now two: 1) can we get (\ref{mu}) relativistically and 2) can we do it without invoking any new degree of freedom rather than the scalar field $\phi$ and the graviton?

The first question is easier. The canonical scalar field action is nothing else than 
\be
{\cal{L}} =-\frac{1}{2}
\sqrt{-g}g^{\alpha\beta}\partial_{\alpha}\phi\partial_{\beta}\phi\ . \label{m2}
\ee 
As we noticed before, the case (\ref{b}) is equivalent to a rescale of time by a factor 
$\sqrt{\mu}$, if $\dot\mu\simeq 0$. Let us extend this rescaling on all coordinates.
 In this case what we want is that, on a DeSitter space in which $H\simeq {\rm const}$,
 $\partial_\alpha\rightarrow \sqrt{\mu}\partial_\alpha=\sqrt{1+3H^2/M^2}\partial_\alpha$. 
In an almost DeSitter Universe we have that $G^{\alpha\beta}\simeq -3H^2 g^{\alpha\beta}$. The covariant
 Lagrangian implementing (\ref{mu}) is then
\be\label{lag}
{\cal{L}} =-\frac{1}{2}\sqrt{-g}\Delta^{\alpha\beta}\partial_\alpha\phi\partial_\beta\phi\ , 
\ee 
where 
\be
\Delta^{\alpha\beta}\equiv g^{\alpha\beta}-\frac{G^{\alpha\beta}}{M^2}\ .
\ee
We then achieved an enhancement of the scalar field friction by a covariant gravitational interaction; for this reason, we have called this mechanism ``gravitationally enhanced friction''.

We are now only left to answer the second question. First of all we note that, because of Bianchi identities ($\nabla_\alpha G^{\alpha\beta}=0$), the field equation for the scalar field is second order. Therefore, no extra propagating degree of freedom appears. One may now wonder about metric variation in the Lagrangian (\ref{lag}). 

In Arnowitt-Deser-Misner (ADM) formalism, a generic metric may be written as \cite{Misner:1974qy, Wald:1984rg}
\be
ds^2=-N^2 dt^2+h_{ij}(N^idt+dx^i)(N^jdt+dx^j)
\ee
and only the spatial metric $h_{ij}$ propagates in GR. There, the lapse and the shift ($N,N^i$), turn out to be just Lagrange multipliers for the Hamiltonian ($G^{tt}$) and momentum constraints. These 4 conditions leave only 2 independent degrees of freedom propagating in $h_{ij}$. The Lagrangian (\ref{lag}) preserves the number of GR constraints and does not introduce higher time derivatives for the propagating fields. This implies that no new degree of freedom is generated with respect to GR coupled minimally with a scalar field if (\ref{lag}) is used \cite{Sushkov:2009hk, Germani:2010gm}. 

To show this, we just need to check that lapse and shift do not have time evolution and that no higher time derivatives of $h_{ij}$ and $\phi$ are generated. The only dangerous term that would produce higher time derivatives and/or propagation of lapse and shift is obviously $\sim G^{tt}\dot\phi^2$. However, since the $G^{tt}$ term is the gravitational Hamiltonian, it only contains one derivative of $h_{ij}$. This implies that no dangerous higher (time) derivative is generated. By diffeomorphism invariance then, any higher derivatives will not be generated (see \cite{Sushkov:2009hk} for explicit calculation). 

The full action of a GEF theory is then 
\be\label{action}
S=\int d^4x\sqrt{-g}\left[\frac{M_p^2}{2}R-\frac{1}{2}\Delta^{\alpha\beta}\partial_{\alpha}\phi\partial_\beta\phi-V\right]\ .
\ee
In a FRW background, the Friedmann and field equations read
\be\label{eq0}
H^2=\frac{1}{3M_p^2}\left[\frac{\dot\phi^2}{2}\left(1+9\frac{H^2}{M^2}\right)+V\right],\quad  \partial_t\left[a^3\dot\phi\left(1+3\frac{H^2}{M^2}\right)\right]=-a^3V'\ .
\ee
During slow roll  in the high friction limit ($H^2/M^2\gg 1$), equations (\ref{eq0}) are simplified as
\be\label{sr}
H^2\simeq\frac{V}{3M_p^2}\ ,\quad \dot\phi\simeq-\frac{V'}{3H}\frac{M^2}{3H^2}\ .
\ee
Consistency of equations (\ref{sr}) requires the slow roll parameters to be small, i.e.
\be\label{srp}
\epsilon\equiv-\frac{\dot H}{H^2}\ll1\ ,\quad \delta\equiv \frac{\ddot\phi}{H\dot\phi}\ll 1 \ .
\ee
By explicit calculation, using (\ref{srp}), one can show that
\be\label{srpv}
\!\!\!\!\!\!\!\!\!\!\!\!\!\!\epsilon\simeq \frac{V'^2M_p^2}{2V^2}\frac{M^2}{3H^2}\simeq\frac{3}{2}\frac{\dot\phi^2}{M^2 M_p^2}\ ,\ 
\delta\simeq -\frac{V'' M_p^2}{V}\frac{M^2}{3H^2}+3\epsilon = -\eta + 3\epsilon \ ,\ 
\eta\equiv \frac{V'' M_p^2}{V}\frac{M^2}{3H^2}\ .
\ee

We see then that, no matter how big are the slow roll parameters in the GR limit ($M/H \rightarrow\infty$), i.e.
\be\label{sr2}
\epsilon_{GR}\equiv\frac{V'^2M_p^2}{2V^2}~~~~~\mbox{and}~~~~~\eta_{GR}\equiv \frac{V'' M_p^2}{V},
\ee
there is always a choice of scale $M^2\ll 3H^2$, during inflation, such that (\ref{srpv}) are small. This is the power of the GEF mechanism.\footnote{Note that GEF is not the only way to modify slow roll parameters. One can modify them by introducing self derivative couplings of the scalar \cite{Nakayama:2010sk, Kamada:2010qe}, nonminimal Ricci scalar coupling \cite{Futamase:1987ua, Bezrukov:2007ep} or both \cite{Lerner:2010mq}. All of them effectively flatten the potential but do not increase {\it friction} meaning that they do not slow down the scalar field by ``dissipating'' into external fields, i.e. the scalar field equation cannot be approximated in the form of Eq.~(\ref{friction}).}

\section{Quadratic action: linear perturbations and spectral index}

In the ADM formalism, the action~(\ref{action}) becomes
\be\label{ADMaction}
S &=& \int d^3xdt \frac{M_p^2}{2}\sqrt{h} \Bigg[  {}^{(3)}R \left(N+ \frac{\dot{\phi}^2}{2N M^2M_p^2}\right) \nonumber\\
&&+  (E_{ij}E^{ij}-E^2) \left(\frac1N - \frac{\dot{\phi}^2}{2N^3 M^2M_p^2}\right) 
+ \frac{\dot{\phi}^2}{NM_p^2}- \frac{2NV}{M_p^2} \Bigg],
\ee 
where the uniform-field gauge has been chosen:
\be
E_{ij} &=& \frac{1}{2}(\dot{h}_{ij} - D_{i}N_j - D_jN_i),\quad E = h^{ij }E_{ij},\\
\delta\phi(x,t)&=& 0,\quad h_{ij} = a^2 e^{2\zeta}(\delta_{ij}+\gamma_{ij} + \frac12 \gamma_{il}\gamma_{lj}),\quad
 D^i \gamma_{ij} = 0, \quad h^{ij}\gamma_{ij} = 0
\ee
to second order.
Varying the action~(\ref{ADMaction}) with respect to $N_j$, one finds the momentum constraint equation
\be\label{momentum_constraint}
D_i \Bigg[\left( \frac{1}{N} -\frac{\dot{\phi}^2}{2N^3M^2M_p^2} \right)(E^i{}_j - h^i{}_j E) \Bigg] = 0.
\ee 
We solve this equation to first order by setting $N = 1+ N_1$, $N_i = \partial_i \psi +N_i^T$ and $h_{ij} = a^2[(1+2\zeta)\delta_{ij} + \gamma_{ij}]$, where $D^i N_i^T = 0$. 
We then find a solution for the lapse function
\be\label{lapse}
N_1 = \frac{\Gamma}{H}\dot{\zeta},\qquad \Gamma \equiv \frac{1-\frac{\dot{\phi}^2}{2M^2M_p^2}}{1- \frac{3\dot{\phi}^2}{2M^2M_p^2}}.
\ee
Varying the action~(\ref{ADMaction}) with respect to $N$, one finds the hamiltonian constraint equation
\be\label{hamiltonian_constraint}
{}^{(3)}R\left( N^2 - \frac{\dot{\phi}^2}{2M^2M_p^2}\right)
- (E_{ij}E^{ij}-E^2)\left(1- \frac{3\dot{\phi}^2}{2N^2M^2M_p^2}\right)
- \frac{\dot{\phi}^2}{M_p^2} - \frac{2N^2V}{M_p^2} = 0.
\ee
This equation gives the Friedmann equation~(\ref{eq0}) to zeroth order. Solving Eq.~(\ref{hamiltonian_constraint}) to first order, one obtains a solution for the shift function
\be\label{shift}
\psi &=& - \frac{\Gamma}{H}\zeta + \chi,\qquad 
\partial_i^2 \chi = \frac{a^2\Sigma}{H^2} \frac{ \Gamma^2 }{ 1-\frac{ \dot{\phi}^2 }{ 2M^2M_p^2 } }\dot{\zeta},\qquad
\Sigma \equiv \frac{\dot{\phi}^2}{2M_p^2} 
\left[1+\frac{3H^2 (1+ \frac{ 3\dot{\phi}^2 }{ 2M^2M_p^2 }) }{ M^2(1-\frac{ \dot{\phi}^2 }{ 2M^2M_p^2 }) }\right],\nonumber\\
N_i^T &=& 0,
\ee
where the Friedmann equation~(\ref{eq0}) and the lapse~(\ref{lapse}) have been used. Note that $\Gamma$ always comes with $H^{-1}$, thus it implies the modified Hubble scale. We have used a notation, $\Sigma$, that resembles one used in general single field inflation \cite{Seery:2005wm, Chen:2006nt}.

Expanding the action~(\ref{ADMaction}) to second order and ignoring tensor modes, we obtain the quadratic action in $\zeta$ after a few integration by parts
\be\label{action2}
S_{\zeta^2} &=& \int d^3xdt\ M_p^2a^3 \Bigg[ \frac{\Gamma ^2\Sigma}{H^2}{\dot \zeta}^2 
-\frac{ \epsilon_s }{a^2}(\partial_i\zeta)^2 \Bigg],\nonumber\\
\epsilon_s &\equiv&  \frac{d}{adt} \left[\frac{a\Gamma}{H}\left( 1-\frac{\dot{\phi}^2}{2M^2M_p^2} \right)  \right]  -\left( 1+\frac{\dot{\phi}^2}{2M^2M_p^2} \right) ,
\ee
where we have used the lapse~(\ref{lapse}), the shift~(\ref{shift}), the Friedmann equation~(\ref{eq0}) and the Raychaudhuri equation
\be
-\frac{\dot{H}}{H^2}\left( 1-\frac{\dot{\phi}^2}{2M^2M_p^2} \right) = \frac{\dot{\phi}^2}{2H^2M_p^2} +\frac32 \frac{\dot{\phi}^2}{M^2M_p^2} - \frac{\ddot\phi\dot{\phi}}{HM^2M_p^2}.
\ee  
Note that all of above expressions recover those of GR in the limit, $M/H \to \infty$.

In the following, we shall always take the high friction limit, $H\gg M$. According to the slow roll equation~(\ref{sr}), $\dot{\phi}$ goes to zero as $M^2$ does so that $\dot\phi^2/(M^2M_p^2) \ll 1$.
Note that in high friction limit, $\Gamma \to 1$, $\Sigma \to [\dot{\phi}^2/(2M_p^2)]\times [3H^2/M^2]$ and $c_s^2\to 1$, the quadratic Lagrangian~(\ref{action2}) becomes
\be
{\cal L}_{\zeta^2}^{\rm GEF} \simeq \frac{3a^3\dot{\phi}^2}{2M^2}\big[\dot{\zeta}^2  - (\partial \zeta)^2\big]
\simeq M_p^2 a^3 \epsilon \big[\dot{\zeta}^2  - (\partial \zeta)^2\big],
\ee
where $(\partial \zeta)^2=(\partial_i \zeta)^2/a^2$.
One can clearly see that the normalization is dependent on the background values.
However, in both GR and GEF limits the spacetime becomes quasi-DeSitter with a small deviation parameterized by $\epsilon$.

In order to quantize fields, one needs to canonically normalize them. We do this by \cite{Mukhanov:2005sc, Garriga:1999vw}\footnote{We could quantize a gauge-invariant variable, $\bar{v} \equiv z(\zeta - H\delta\phi/\dot\phi)$, instead of $v$.}
\be\label{action2_norm}
S_{\zeta^2} = \int d^3xd\tau \frac12\left[ v'^2 - c_s^2 (\partial_i v)^2 + \frac{z''}{z}v^2 \right],\nonumber\\
v = z \zeta, \qquad 
z = a \frac{M_p \Gamma}{H}\sqrt{2\Sigma} ,\qquad
c_s^2 = \frac{ H^2\epsilon_s }{ \Gamma^2 \Sigma},
\ee
where $\tau$ is the conformal time and the prime denotes the derivative with respect to $\tau$. 
Note that we have integrated by parts to get the time dependent mass term. Note also that $\Sigma$, $\dot{\phi}$ and $H$ are slowly changing variables during inflation.

From the Friedmann equation~(\ref{eq0}), we have  $3\dot\phi^2/(2M^2M_p^2) \le 1$; thus the sound speed squared is positive definite and sub-luminal, $0< c_s^2 <1$, i.e. there is no tachyonic propagation.\footnote{In the high friction limit, $c_s^2 \simeq 1 - 4\dot\phi^2/(M^2M_p^2) \simeq 1- 8\epsilon/3$ and $\epsilon \simeq 3\dot\phi^2/(2M^2M_p^2)\ll 1$.} 
Moreover, $\Gamma^2 \Sigma/H^2 >0$ in the action~(\ref{action2}) indicates that the curvature perturbations cannot be ghost-like in the FRW background.

One obtains the Mukhanov-Sasaki equation by varying the action~(\ref{action2_norm}) with respect to $v$. In the Fourier space,
\be\label{eq:mukhanov-sasaki}
v_k'' + \left( c_s^2 k^2 - \frac{z''}{z}\right)v_k = 0,
\ee
where  $a\simeq -1/(H\tau)$ and $z''/z \simeq 2/\tau^2$ in the quasi-DeSitter background. 
We have defined the mode function $v_k$ by promoting $v$ to an operator $\hat{v}$ as
\be
\hat{v}(\tau,{\bf x}) &=& \int \frac{d^3k}{(2\pi)^3}\hat{v}(\tau, {\bf k}) e^{i {\bf k}\cdot {\bf x}},\quad
\hat{v}(\tau,{\bf k}) = v_k \hat{a}({\bf k}) + v_{-k}^* \hat{a}^{\dagger}(-{\bf k}),\nonumber\\
\big[\hat{a}({\bf k}), \hat{a}^{\dagger}({\bf k'})\big] &=& (2\pi)^3\delta^3({\bf k}- {\bf k'}),\quad
\big[\hat{a}({\bf k}), \hat{a}({\bf k'})\big] =\big[\hat{a}^{\dagger}({\bf k}), \hat{a}^{\dagger}({\bf k'})\big] = 0, 
\ee
where $v_k$ and $v_k^*$ are two independent solutions that obey the normalization condition:
\be
v'_k v_k^* -v_k {v_k^*}' = -i.
\ee
The vacuum state $|0\rangle$ is defined by
\be
\hat{a}({\bf k})|0\rangle = 0.
\ee

Normalizing by the standard Bunch-Davis vacuum in the asymptotic past, $v(k\tau \to -\infty) = e^{-ic_s k\tau}/\sqrt{2c_s k}$, the solution of Eq.~(\ref{eq:mukhanov-sasaki}) is thus given by
\be
\zeta_k = \frac{v_k}{z} = \frac{-i  e^{-ic_sk\tau}}{z\sqrt{2}(c_s k)^{3/2}\tau}(1+ic_sk\tau)
\simeq  \frac{i H e^{-ic_sk\tau}}{2\sqrt{\epsilon_s c_s}k^{3/2}M_p}(1+ic_sk\tau).
\ee
The power spectrum of $\zeta$ is defined by the two-point correlation function:
\be
\langle \hat{\zeta}(\tau,{\bf k})\hat{\zeta}(\tau,{\bf k'}) \rangle = (2\pi)^3\delta^3({\bf k}+{\bf k'})P_{\zeta}(k),\quad
P_{\zeta}(k) \equiv |\zeta_k|^2 \simeq \frac{H^2}{4k^3\epsilon_sc_sM_p^2}.
\ee
Equivalently, the dimensionless power spectrum is given by
\be\label{ps}
{\cal P}_{\zeta} \equiv \frac{k^3}{2\pi^2}|\zeta_k|^2
\simeq \frac{H^2}{8\pi^2\epsilon_s c_s M_p^2 }. 
\ee

If we match the spectrum in the high friction limit~(\ref{ps}) with the WMAP data \cite{Komatsu:2010fb},
\be
{\cal P}_{\zeta} = 2\times 10^{-9},
\ee
we get a relation
\be\label{relation_ps}
\frac{M^2}{H^2} = \frac{10^9}{8\pi^2}\frac{V^3}{M_p^6V'^2}.
\ee

The spectral tilt of Eq.~(\ref{ps}) is given by 
\be\label{tilt}
n_s - 1 &\equiv& \left.\frac{d \ln {\cal P}_{\zeta}}{d\ln k}\right|_{c_sk=aH} 
\approx -2\epsilon - \frac{\dot{\epsilon_s}}{\epsilon_s H} - \frac{\dot{c_s}}{c_sH}
= -2\epsilon -2\delta +{\cal O}(\epsilon^2)\nonumber\\
&\simeq& \frac{M^2}{H^2}M_p^2\Bigg[-\frac{4}{3}\frac{V'^2}{V^2}+\frac23\frac{V''}{V} \Bigg]
= -8\epsilon +2\eta,
\ee
where we have used $d\ln k \approx d\ln a$ and the slow-roll equations in the high friction limit.\footnote{We have used relations
\be
\epsilon = \frac32 \frac{\dot\phi^2}{M^2M_p^2},\
\frac{\dot\epsilon}{\epsilon H} = 2\delta,\ 
\frac{\dot\epsilon_s}{\epsilon_s H} = 2\delta,\ 
\frac{\dot\Gamma}{\Gamma H} = \frac43 \epsilon\delta,\ 
\frac{\dot\Sigma}{\Sigma H} = 2\delta - 2\epsilon,\ 
\frac{\dot{c_s}}{c_sH} = {\cal O}(\epsilon^2)
\ee
to the leading order in slow roll. Note that $\epsilon_s \simeq \epsilon -(5/9)\epsilon^2+(2/3)\epsilon\delta$ in the high friction limit.}
Note that the relation is different from the standard one, $n_s-1 = -4\epsilon - 2\delta= -6\epsilon +2\eta$, in the GR limit \cite{Liddle:2000cg}.
Given a shape of a potential, one can constrain a model by using the relations~(\ref{relation_ps}) and (\ref{tilt}).

The running of the spectral index~(\ref{tilt}) is given by
\be
\left.\frac{d n_s}{d\ln k}\right|_{c_sk = aH} = -6 \epsilon \delta - 2 \delta\delta' + 2\delta^2,
\ee
where $\delta' \equiv \dddot\phi/(\ddot\phi H)$.

\section{Tensor to scalar ratio}

In the quadratic action, scalar and tensor modes are decoupled. Expanding the action~(\ref{ADMaction}) to second order, we also obtain the quadratic action in $\gamma_{ij}$ after integration by parts
\be\label{gg}
S_{\gamma^2} = \int d^3xdt\ \frac{M_p^2}{8}a^3\Bigg[\left(1-\frac{\dot\phi^2}{2M^2M_p^2}\right)\dot{\gamma}_{ij}^2 -  \left(1+\frac{\dot\phi^2}{2M^2M_p^2}\right)\frac{1}{a^2}(\partial_k\gamma_{ij})^2\Bigg].
\ee
 In order to quantize gravitons, we canonically normalize by
 \be
 S_{\gamma^2} &=& \sum_{\lambda=\pm 2}\int d^3x d\tau\  \frac12 \Bigg[ v_{t}'^2 - c_t^2 (\partial_i v_t)^2 +\frac{z_t''}{z_t}v_t^2 \Bigg],\nonumber\\
 v_t &=& z_t \gamma_{\lambda},\quad 
 z_t = aM_p \frac{ \sqrt{e_{ij}^{\lambda}e_{ij}^{\lambda}} }{2} \sqrt{ 1-\frac{\dot{\phi}^2}{2M^2M_p^2} }, \quad
 c_t^2 = \frac{1+\frac{\dot{\phi}^2}{2M^2M_p^2}  }{1-\frac{\dot{\phi}^2}{2M^2M_p^2}} ,
 \ee
 where $\gamma_{ij} = \gamma_{+}e_{ij}^+ + \gamma_{-}e_{ij}^{-}$ is quantized to each helicity mode \cite{Weinberg:2008zzc}.
 Conventionally, the polarization tensor is normalized to $e_{ij}^{\lambda}e_{ij}^{{\lambda}'}=2\delta_{\lambda{\lambda}'} $, but we keep it unspecified here.
 Note that $c_t^2 > 1$ does not mean "super-luminal" because the causal structure is set by the propagation of gravitational waves \cite{Germani:2010ux}.
 
 The tensor modes also obey the Mukhanov-Sasaki equation~(\ref{eq:mukhanov-sasaki}) with $v_t,\ z_t$ and $c_t$. 
 The mode function is given by 
 \be
 \gamma_{\lambda}(k) = \frac{-ie^{-ic_t k\tau}}{z_t\sqrt2(c_tk)^{3/2}\tau}(1+ic_t k\tau) 
 \simeq \sqrt{\frac{2}{e_{ij}^{\lambda}e_{ij}^{\lambda}}} \frac{ iHe^{-ic_t k\tau}}{ \sqrt{1+\frac{\dot\phi^2}{2M^2M_p^2}}\sqrt{c_t}k^{3/2}M_p}(1+ic_t k\tau),
 \ee
 where we have normalized by the Bunch-Davis vacuum.
 The dimensionless power spectrum of gravitational waves is given by
\be\label{ps_t}
{\cal P}_{\gamma} = \frac{k^3}{2\pi^2}\sum_{\lambda=\pm 2}|\gamma_{\lambda}(k)e_{ij}^{\lambda}(k)|^2
\simeq \frac{2H^2}{\pi^2c_t M_p^2\left(1+\frac{\dot\phi^2}{2M^2M_p^2}\right)},
\ee
where we have assumed ${\cal P}_{\gamma+} = {\cal P}_{\gamma-} ={\cal P}_{\gamma}/2$. Both helicity states are statistically independent with the same amplitude unless there are parity-violating interactions, such as $\phi F\tilde{F}/f$ and $\phi R\tilde{R}/f$ \cite{Lue:1998mq, Saito:2007kt, Gluscevic:2010vv, Sorbo:2011rz}.
In the UV-protected inflation, we have such interactions and will come back to this point in Sec.~\ref{sec:barnaby-peloso}.
 
The spectral index of Eq.~(\ref{ps_t}) is given by
\be
n_t \equiv \left.\frac{d\ln {\cal P}_{\gamma} }{d\ln k} \right|_{c_t k=aH} 
\approx -2\epsilon - \frac{\dot{c_t}}{c_t H} - \frac{ \ddot\phi\dot\phi/(M^2M_p^2) }{ \Big( 1+\frac{\dot\phi^2}{2M^2M_p^2} \Big)H }
= -2\epsilon +{\cal O}(\epsilon^2),
\ee
 where we have used $d\ln k \approx d\ln a$. Regardless of the potential shape, the gravitational wave spectrum must be red-tilted.
 
 The ratio of tensor to scalar spectrum is given by
 \be\label{ratio}
 r \equiv \frac{{\cal P}_{\gamma}}{{\cal P}_{\zeta}} = \frac{16\epsilon_s c_s}{c_t  \Big( 1+\frac{\dot\phi^2}{2M^2M_p^2} \Big)}
 = 16\epsilon +{\cal O}(\epsilon^2)
 \ee
 whose definition agrees with that of the WMAP team (see Sec.~3.2 of \cite{Komatsu:2008hk}). 
 Therefore, we get the consistency relation between $r$ and $n_t$
 \be
 r = -8 n_t,
 \ee
 which is the same as that of GR to first order \cite{Liddle:2000cg}.
 
 The Lyth bound tells us that detectable gravitational waves require super-Planckian field variation, $\Delta\phi \gtrsim 2\ {\rm to}\ 6M_p$ \cite{Lyth:1996im, Lyth:1998xn}.
 
Under GEF, this bound reads
 \be\label{lyth}
 \left( \frac{r}{0.1} \right)^{1/2} \lesssim \frac{H}{20 M} \frac{\Delta \phi}{M_p}\frac{50}{N_e},
 \ee
 where $N_e$ is the number of efolds given by $N_e = \int  Hdt$.

 Although at first sight the bound (\ref{lyth}) seems to allow detectable gravitational waves for sub-Planckian values of the field, in fact, it actually requires the canonically normalized inflaton [$\tilde\phi\sim (H/M)\phi$] to be super-Planckian. In this sense, the Lyth bound is not modified.

\section{Cubic action: gauge transformation and strong coupling scales}

In order to compute the leading order scattering amplitude and non-Gaussianity, the cubic terms of the action are needed. 
In principle, one can expand the action~(\ref{ADMaction}) to third order, but it requires a lot of integration by parts to reduce the form simple enough. 
We find it more convenient to get them in the flat gauge, 
\be\label{metric_flat}
\delta\phi \equiv \pi(x,t),\quad
h_{ij} = a^2 (\delta_{ij} +\gamma_{ij}+\frac12 \gamma_{il}\gamma_{lj}),\quad
D^i\gamma_{ij}=0,\quad
h^{ij}\gamma_{ij}=0.
\ee
Although the ADM action is not as simple as that in the uniform-field gauge~(\ref{ADMaction}), one can get the cubic action in the flat gauge as follows.

The constraints are needed only to first order for the cubic action, and we have already solved them in the uniform-field gauge. We make a time reparametrization from uniform-field slicing to flat slicing \cite{Maldacena:2002vr},
\be
{\tilde t} = t + T,\qquad
T = -\frac{\pi(x, \tilde{t})}{\dot\phi(t)},
\ee
where we have found $T$ by Taylor expanding the relation
\be\label{phi-slice}
\phi(t+T)+\pi(x,t+T) = \phi(t)
\ee
to first order.
Since 
\be
h_{ij}^{\pi}(x,t+T) dx^i dx^j= h_{ij}(x,t) dx^i dx^j
\ee
to first order, $\zeta$ and $\gamma_{ij}$ are transformed as
\be\label{transform_zeta}
\zeta(x,t) = HT = -H\frac{\pi(x,\tilde{t} )}{\dot\phi(t)},\quad
\gamma_{ij}(x, t)=\gamma_{ij}^{\pi}(x,\tilde{t}) ,
\ee
where we use the label $\pi$ to indicate flat gauge quantities.

It is then clear that tensor modes are invariant under the time reparametrization.
Since $d\tilde{t}=dt + dT = dt + \dot{T}dt + \partial_i T dx^i$ and
\be
-N_{\pi}^2(x,t+T)d\tilde{t}^2 + h_{ij}^{\pi}(x,t+T)\left( N_\pi^i(x,t+T)dx^j d\tilde{t}+N_\pi^j(x,t+T)dx^i d\tilde{t} \right)\qquad \nonumber\\
= -N^{2}(x,t)dt^2 + h_{ij}(x,t)\left( N^i(x,t)dx^j dt+N^j(x,t)dx^i dt \right)
\ee
to first order, the lapse and shift functions are transformed as
\be\label{constraints_flat}
N_1^{\pi}(x,\tilde{t}) &=& N_1(x,t)-\dot{T}
= -\Gamma\frac{\dot H}{H}\frac{\pi}{\dot \phi} + (1- \Gamma)\frac{d}{dt}\left(\frac{\pi}{\dot \phi}\right),\\
N_{\pi}^i(x,\tilde{t}) &=& N^i(x,t) + \partial^i T
=\partial^i \psi_{\pi}, \quad
\psi_{\pi} = (\Gamma - 1)\frac{\pi}{\dot\phi}+\chi, \nonumber\\
\chi &=& \frac{a^2\Sigma}{H^2}\frac{\Gamma^2}{1-\frac{\dot\phi^2}{2M^2M_p^2}}\partial_i^{-2}\frac{d}{dt}\left(-H\frac{\pi}{\dot\phi}\right),
\ee
respectively. Here $\chi$ has the same form as in Eq.~(\ref{shift}). Note that the limit of $\dot\phi^2/(M^2M_p^2)\to 0$ and $\Gamma\to 1$ reproduces the corresponding expressions in GR.

Plugging the metric~(\ref{metric_flat}) and constraints~(\ref{constraints_flat}) in the action~(\ref{action}) and taking the high friction limit, we obtain the quadratic and cubic Lagrangians
\be
{\cal L}_{\pi^2}^{GEF}&\simeq& \frac{3H^2}{2M^2}a^3\Big[ \dot\pi^2 - (\partial \pi)^2 \Big],\\
{\cal L}_{\pi^3}^{GEF}&\simeq& \frac{H^4}{M^4}a^3 \Big[C_1 \pi {\dot \pi}^2 +C_2  \dot\pi^3 +C_3  \partial^2\pi \dot\pi^2 +C_4\pi(\partial \pi)^2 + C_5  \dot{\pi}(\partial \pi)^2 \nonumber\\
&&\qquad + C_6 \dot\pi\partial^i \pi\partial_i\chi + C_7 \dot\pi^2\partial^2\chi
\Big],\\
C_1 &=& -\frac{27}{4}\frac{\dot\phi}{HM_p^2},\ 
C_2 = \frac92 \frac{\dot\phi}{H^2M_p^2},\ 
C_3 = -3\frac{\dot\phi}{H^3M_p^2}, \ 
C_4 = \frac94 \frac{\dot\phi}{HM_p^2},\ 
C_5 = -5\frac{\dot\phi}{H^2M_p^2},\nonumber\\
C_6&=& -3,\ C_7= -\frac1H, \nonumber
\ee
where several integration by parts have been done and higher orders in slow roll have been ignored.
Note that the high friction limit automatically guarantees slow rolling.

Finally, we transform back to the uniform-field gauge where the curvature perturbations are conserved outside the horizon.\footnote{This procedure is valid only in the lowest order of slow roll since the redefined $\zeta$ may not be conserved in general. We thank Misao Sasaki for pointing this out.} 
Expanding Eq.~(\ref{phi-slice}) to second order, one gets
\be
T_2 = -\frac{\pi}{\dot\phi} -\frac12\frac{\ddot\phi\pi^2}{\dot\phi^3} + \frac{\dot\pi\pi}{\dot\phi^2}.
\ee
Using this second order gauge transformation,
\be\label{3scalar}
{\cal L}_{\pi^2}^{GEF}&\simeq& M_p^2a^3\epsilon\Big[ \dot\zeta^2 - (\partial \zeta)^2 \Big] + {\cal L}_{\zeta^3}^{\rm redef},\\
{\cal L}_{\pi^3}^{GEF}&\simeq& a^3 \Big[c_1 \zeta {\dot \zeta}^2 +c_2  \dot\zeta^3 +c_3  \partial^2\zeta \dot\zeta^2 +c_4\zeta(\partial \zeta)^2 + c_5  \dot{\zeta}(\partial \zeta)^2 \nonumber\\
&&\qquad + c_6 \dot\zeta\partial^i \zeta\partial_i\chi + c_7 \dot\zeta^2\partial^2\chi
\Big],\nonumber\\
c_1 &=& \frac{27}{4}\frac{\dot\phi^4}{M^4M_p^2}=3M_p^2\epsilon^2,\ 
c_2 = -\frac92 \frac{\dot\phi^4}{HM^4M_p^2} = -2\frac{M_p^2}{H}\epsilon^2,\ 
c_3 = 3\frac{\dot\phi^4}{H^2M^4M_p^2} = \frac43\frac{M_p^2}{H^2}\epsilon^2, \nonumber\\ 
c_4 &=& -\frac94 \frac{\dot\phi^4}{M^4M_p^2}=-M_p^2\epsilon^2,\ 
c_5 = 5\frac{\dot\phi^4}{HM^4M_p^2} = \frac{20}{9}\frac{M_p^2}{H}\epsilon^2,\nonumber\\
c_6&=& -3\frac{\dot\phi^2H^2}{M^4},\ c_7= -\frac{\dot\phi^2H}{M^4}. \nonumber
\ee
Reorganizing the terms in order of $\zeta$,
\be
{\cal L}_{\zeta^3} = {\cal L}_{\pi^3} + {\cal L}_{\zeta^3}^{\rm redef},
\ee
 where ${\cal L}_{\zeta^3}^{\rm redef} $ is given by the field redefinition, $\zeta \to \zeta + (\epsilon/2 +\delta/2)\zeta^2$, on super-horizon scales \cite{Maldacena:2002vr}. 
 The field redefinition can also be obtained by the so-called $\delta N$ formalism \cite{Sasaki:1995aw}:
\be
 \zeta &=&
 N_{\phi}\pi + \frac12N_{\phi\phi}\pi^2,\ 
 N_{\phi} = -\frac{H}{\dot\phi},\ 
 N_{\phi\phi} = -\frac{\dot H}{\dot\phi^2}+\frac{H\ddot\phi}{\dot\phi^3},\nonumber\\
 \zeta &\to& \zeta + \frac12\left(-\frac{\dot H}{H^2}+\frac{\ddot \phi}{H\dot\phi} \right)\zeta^2.
\ee
 
 \subsection{Strong coupling scales}

As discussed before, in the GEF theories of inflation, a new scale $M$ and, tight to that, a new non-renormalizable interaction, are introduced. One may then be tempted to associate $M$ or better the scale $\Lambda_{flat}=(M^2 M_p)^{1/3}$ (obtained by expanding (\ref{action}) around the Minkowski background) to the strong coupling scale of the graviton-inflaton system. However, this naive expectation is wrong in a non-trivial background. This is due to the fact that, in a non-trivial background, the inflaton $\phi$ {\it is not} canonically normalized due to the background value of the Einstein tensor in the kinetic interaction $G^{\alpha\beta}\partial_\alpha\phi\partial_\beta\phi$. Moreover, in a non-trivial background, graviton and scalar field kinetic terms mix and therefore, in order to obtain the correct {\it perturbative} strong coupling scale, diagonalization of the kinetic terms must be performed before.

In a single field inflation with GEF, fortunately, this complicated process have a simple shortcut.

During inflation, one can indeed automatically diagonalize the kinetic terms by considering the gauge $\delta\phi=0$ \cite{Germani:2010hd}. In this gauge, the canonical normalization of the graviton is shifted as [see Eq.~(\ref{gg})]
\be
\frac{M_p}{\sqrt2}\rightarrow \frac{M_p}{\sqrt2}\Bigg(1-\frac{\dot\phi^2}{2M^2 M_p^2}\Bigg)^{1/2}\simeq \frac{M_p}{\sqrt2}\ ,
\ee
where the last equality has been obtained by noticing that $\epsilon\simeq 3\dot\phi^2/(2M_p^2 M^2) \ll 1$ from Eq.~(\ref{sr}).

At the end of the previous section, we have showed that during GEF inflation, scattering vertices involving three-scalars are suppressed by a scale $\Lambda\simeq M_p/\sqrt{\epsilon}\gg M_p $ [see Eq. (\ref{3scalar}) and canonically normalize scalars]. As in the canonical GR case, one can then show that this $1/\sqrt{\epsilon}$ enhancement of the strong coupling scale is also true each time a scalar is involved in a scattering process during inflation. Thus, as in GR, in single field GEF inflation the strong coupling scale of the system is determined by graviton only interactions and therefore can be identified with $\sim M_p$.

Note that this property is drastically modified in the multi-field scenarios, like in the New Higgs inflation of \cite{Germani:2010gm}. There, the non-inflating scalar introduces a much lower strong coupling scale that, during inflation, is $\Lambda\simeq (H^2 M_p)^{1/3}\ll M_p$ \cite{Germani:2010gm}; nevertheless this model is still weakly coupled.

Let us finally discuss the quantum gravity scale. The graviton is supposedly universally coupled with any form of matter; therefore, for particles minimally coupled to gravity but not interacting with $\phi$, we can in principle perform scattering experiments such to probe up to $M_p$ {\it independently} on the background. In this respect then, $M_p$ is the reference scale of gravity. For this reason, we will always require that all particle masses should be below $M_p$ \cite{Dvali:2010ue}.
   
\section{Non-Gaussianities in single field GEF}

We compute the non-Gaussian feature of the scalar fluctuations. We use the uniform-field gauge variable, $\zeta$, since it is conserved outside  the horizon (at least at order $\epsilon$).
The leading order effect appears in the bispectrum or the three-point function. 
Since ${\cal L}^{GEF}_{\zeta^3} \sim {\cal O}(\epsilon^2)$, we get $f_{NL} \sim {\cal O}(\epsilon)$.

As in the power spectrum, the bispectrum of $\zeta$ is defined by the three-point correlation function:
\be
\langle \hat\zeta(\tau,{\bf k}_1)\hat\zeta(\tau,{\bf k}_2)\hat\zeta(\tau,{\bf k}_3)\rangle \equiv (2\pi)^3\delta^3({\bf k}_1+{\bf k}_2 +{\bf k}_3)B_{\zeta}(k_1,k_2,k_3).
\ee 

One can evaluate the three-point correlator by using the in-in formalism \cite{Maldacena:2002vr, Weinberg:2005vy, Koyama:2010xj}.
In the lowest order,
\be\label{in-in}
\langle \hat{\zeta}(0,{\bf k}_1)\hat{\zeta}(0,{\bf k}_2)\hat{\zeta}(0,{\bf k}_3) \rangle = -i\int_{-\infty}^0 d\tau a \langle 0| [\hat{\zeta}(0,{\bf k}_1)\hat\zeta(0,{\bf k}_2)\hat\zeta(0,{\bf k}_3), \hat{H}_{int}(\tau)]|0 \rangle,
\ee
where we have set the initial and final times as $\tau_i = - \infty$ and $\tau_f = 0$, respectively.
The interaction Hamiltonian is given by
\be
\hat{H}_{int}(\tau) = -\int d^3x \hat{\cal L}_{\zeta^3}^{GEF}.
\ee
If there is no interaction, the three-point correlator vanishes as one can see from Eq.~(\ref{in-in}).
For simplicity, we shall suppress the carets ( $\hat{}$ ) on variables, but they should be understood as operators in the following.

The three-point function can be calculated from each term of the interaction Hamiltonian as in \cite{Seery:2005wm, Chen:2006nt, DeFelice:2011zh}:
\be\label{three-point-1}
H_{int}^{(1)}(\tau)&=& -c_1a^3\int d^3x \zeta\dot\zeta^2 \\
&=& - c_1a\int\frac{d^3k_4d^3k_5d^3k_6}{(2\pi)^6}\delta^{3}({\bf k}_4+{\bf k}_5+{\bf k}_6)\zeta(\tau,k_4)\zeta'(\tau,k_5)\zeta'(\tau,k_6),\nonumber\\
\langle \zeta(k_1)\zeta(k_2)\zeta(k_3) \rangle^{(1)} 
&=& (2\pi)^3\delta^{3}({\bf k}_1+{\bf k}_2+{\bf k}_3)\frac{c_1 H^4}{16\epsilon_s^3 M_p^6}\frac{1}{(k_1k_2 k_3)^3}\left(\frac{k_2^2k_3^2}{K} +\frac{k_1k_2^2k_3^2}{K^2}+{\rm sym}\right), \nonumber
\ee
where $k = |{\bf k}|$, $K = k_1+k_2+k_3$ and "sym" denotes the symmetric terms with respect to $k_1,\ k_2,\ k_3$.
Here we have used the canonical commutation relations $\langle0|[a({\bf k}),a^{\dagger}({\bf k}')]|0\rangle=(2\pi)^3\delta^3({\bf k}-{\bf k}')$, and their non-vanishing combinations
\be
&&\langle0| a({\bf k}_1)a({\bf k}_2)a({\bf k}_3)a^{\dagger}(-{\bf k}_4)a^{\dagger}(-{\bf k}_5)a^{\dagger}(-{\bf k}_6) |0\rangle\nonumber\\
&=&\langle0| a({\bf k}_4)a({\bf k}_5)a({\bf k}_6)a^{\dagger}(-{\bf k}_1)a^{\dagger}(-{\bf k}_2)a^{\dagger}(-{\bf k}_3) |0\rangle \nonumber\\
&=&(2\pi)^9\Big[\delta^3({\bf k}_1+{\bf k}_4)\big[ \delta^3({\bf k}_2+{\bf k}_5)\delta^3({\bf k}_3+{\bf k}_6)+\delta^3({\bf k}_2+{\bf k}_6)\delta^3({\bf k}_3+{\bf k}_5) \big]\nonumber\\
&&\qquad+\delta^3({\bf k}_1+{\bf k}_5)\big[ \delta^3({\bf k}_2+{\bf k}_4)\delta^3({\bf k}_3+{\bf k}_6)+\delta^3({\bf k}_2+{\bf k}_6)\delta^3({\bf k}_3+{\bf k}_4) \big]\nonumber\\
&&\qquad+\delta^3({\bf k}_1+{\bf k}_6)\big[ \delta^3({\bf k}_2+{\bf k}_4)\delta^3({\bf k}_3+{\bf k}_5)+\delta^3({\bf k}_2+{\bf k}_5)\delta^3({\bf k}_3+{\bf k}_4) \big]\Big].
\ee
The conformal time integral gives $\int^0_{-\infty(1-i\varepsilon)}d\tau e^{ic_sK\tau}(1-ic_sk_1\tau)=1/(ic_sK)+k_1/(ic_sK^2)$.
Similarly, we find
\be\label{three-point-2}
H_{int}^{(2)}(\tau)&=& -c_2 a^3\int d^3x \dot\zeta^3 \\
&=& - c_2 \int\frac{d^3k_4d^3k_5d^3k_6}{(2\pi)^6}\delta^{3}({\bf k}_4+{\bf k}_5+{\bf k}_6)\zeta'(\tau,k_4)\zeta'(\tau,k_5)\zeta'(\tau,k_6),\nonumber\\
\langle \zeta(k_1)\zeta(k_2)\zeta(k_3) \rangle^{(2)} 
&=& (2\pi)^3\delta^{3}({\bf k}_1+{\bf k}_2+{\bf k}_3)\frac{3c_2 H^5}{8\epsilon_s^3 M_p^6}\frac{1}{k_1k_2 k_3 K^3},\nonumber
\ee
where we have used $\int_{-\infty(1-i\varepsilon)}^0 \tau^2e^{ic_sK\tau}=2i/(c_s^3K^3)$. We have other contributions:
\be\label{three-point-3}
H_{int}^{(3)}(\tau)&=& -c_3a\int d^3x \partial_i^2\zeta \dot\zeta^2 \\
&=& - \frac{c_3}{a} \int\frac{d^3k_4d^3k_5d^3k_6}{(2\pi)^6}\delta^{3}({\bf k}_4+{\bf k}_5+{\bf k}_6)k_4^2\zeta(\tau,k_4)\zeta'(\tau,k_5)\zeta'(\tau,k_6),\nonumber\\
\langle \zeta(k_1)\zeta(k_2)\zeta(k_3) \rangle^{(3)} 
&=& (2\pi)^3\delta^{3}({\bf k}_1+{\bf k}_2+{\bf k}_3)\frac{3c_3 H^6}{4\epsilon_s^3 c_s^2 M_p^6}\frac{1}{k_1k_2 k_3 K^3},\nonumber
\ee
\be\label{three-point-4}
H_{int}^{(4)}(\tau)&=& -c_4a\int d^3x \zeta(\partial_i\zeta)^2 \\
&=& - c_4a\int\frac{d^3k_4d^3k_5d^3k_6}{(2\pi)^6}\delta^{3}({\bf k}_4+{\bf k}_5+{\bf k}_6)({\bf k}_5\cdot{\bf k}_6)\zeta(\tau,k_4)\zeta(\tau,k_5)\zeta(\tau,k_6),\nonumber\\
\langle \zeta(k_1)\zeta(k_2)\zeta(k_3) \rangle^{(4)} 
&=& (2\pi)^3\delta^{3}({\bf k}_1+{\bf k}_2+{\bf k}_3)\frac{c_4 H^4}{16\epsilon_s^3 c_s^2 M_p^6} \frac{1}{(k_1k_2 k_3)^3}\nonumber\\
&\times&\left[ ({\bf k}_1\cdot{\bf k}_2 +{\bf k}_2\cdot{\bf k}_3 +{\bf k}_3\cdot{\bf k}_1 )\Big(-K+ \frac{k_1k_2+k_2k_3+k_3k_1}{K}+\frac{k_1k_2k_3}{K^2}\Big) \right], \nonumber\\
H_{int}^{(5)}(\tau)&=& -c_5a\int d^3x \dot\zeta(\partial_i\zeta)^2 \\
&=& - c_5\int\frac{d^3k_4d^3k_5d^3k_6}{(2\pi)^6}\delta^{3}({\bf k}_4+{\bf k}_5+{\bf k}_6)({\bf k}_5\cdot{\bf k}_6)\zeta'(\tau,k_4)\zeta(\tau,k_5)\zeta(\tau,k_6),\nonumber\\
\langle \zeta(k_1)\zeta(k_2)\zeta(k_3) \rangle^{(5)} 
&=& (2\pi)^3\delta^{3}({\bf k}_1+{\bf k}_2+{\bf k}_3) \frac{c_5 H^5}{32\epsilon_s^3c_s^2 M_p^6}\frac{1}{(k_1k_2k_3)^3}\nonumber\\
&&\times \Bigg[\frac{k_1^2({\bf k}_2\cdot{\bf k}_3)}{K}\Big( 1+\frac{k_2+k_3}{K}+\frac{2k_2k_3}{K^2}\Big)+{\rm sym} \Bigg].\nonumber
\ee

Also, for the terms involving non-local function, $\chi$,
\be
H_{int}^{(6)}(\tau)&=& -c_6 a\int d^3x \dot\zeta\partial_i\zeta\partial_i\chi \\
&=& - c_6\int\frac{d^3k_4d^3k_5d^3k_6}{(2\pi)^6}\delta^{3}({\bf k}_4+{\bf k}_5+{\bf k}_6)({\bf k}_5\cdot{\bf k}_6)\zeta'(\tau,k_4)\zeta(\tau,k_5)\chi(\tau,k_6),\nonumber\\
\langle \zeta(k_1)\zeta(k_2)\zeta(k_3) \rangle^{(6)} 
&=& (2\pi)^3\delta^{3}({\bf k}_1+{\bf k}_2+{\bf k}_3)\frac{c_6H^4}{32\epsilon_s^2c_s^2 M_p^6}\frac{1}{(k_1k_2k_3)^3}\nonumber\\
&&\times\left[ \frac{({\bf k}_1\cdot {\bf k}_2)k_3^2}{K}\Big(2+\frac{k_1+k_2}{K}\Big) +{\rm sym} \right], \nonumber\\
H_{int}^{(7)}(\tau)&=& -c_7a\int d^3x \dot\zeta^2\partial_i^2\chi 
=-\tilde{c}_7a^3\int d^3x\dot\zeta^3\\
&=& - \tilde{c}_7\int\frac{d^3k_4d^3k_5d^3k_6}{(2\pi)^6}\delta^{3}({\bf k}_4+{\bf k}_5+{\bf k}_6)\zeta'(\tau,k_4)\zeta'(\tau,k_5)\zeta'(\tau,k_6),\nonumber\\
\langle \zeta(k_1)\zeta(k_2)\zeta(k_3) \rangle^{(7)} 
&=& (2\pi)^3\delta^{3}({\bf k}_1+{\bf k}_2+{\bf k}_3)\frac{3\tilde{c}_7 H^5}{8\epsilon_s^3 M_p^6}\frac{1}{k_1k_2 k_3 K^3}, \nonumber
\ee
which is the same form as $\dot\zeta^3$ interaction~(\ref{three-point-2}). Coefficients $c_i$'s are given in Eq.~(\ref{3scalar}) and $\tilde{c}_7 \equiv \epsilon c_7 $.

By using the Wick's theorem, we obtain the contribution from field redefinition $\zeta\to \zeta+(\epsilon/2+\delta/2)\zeta^2$:
\be\label{three-point-redef}
B_{\zeta}^{\rm redef}(k_1,k_2,k_3)= \frac{(\epsilon+\delta)H^4}{16\epsilon_s^2c_s^2M_p^4}
\Bigg(\frac{1}{k_1^3k_2^3}+ \frac{1}{k_2^3k_3^3}+\frac{1}{k_3^3k_1^3}\Bigg).
\ee

\subsection{Local form and the consistency relation}\label{sub:local}

Maldacena \cite{Maldacena:2002vr} has noticed that the squeezed limit of the bispectrum is given by 
\be\label{consistency_fnl}
\frac{12}{5}f_{NL}^{\rm local} = \frac{B_{\zeta}(k_1,k_2\to k_1,k_3\to 0)}{P_{\zeta}(k_1)P_{\zeta}(k_3)}= 1-n_s,
\ee
where $f_{NL}^{\rm local}$ is the definition of \cite{Komatsu:2001rj}.
This consistency relation applies to any single-field inflation model \cite{Creminelli:2004yq}. 
We show that it also applies to any single-field model with GEF. In the squeezed limit, we find from Eqs.~(\ref{three-point-1}), (\ref{three-point-4}) and (\ref{three-point-redef})
\be
B_{\zeta}^{(1)}(k_1,k_2\to k_1,k_3\to 0) &=& \frac{3\epsilon}{2}P_{\zeta}(k_1)P_{\zeta}(k_3),\\
B_{\zeta}^{(4)}(k_1,k_2\to k_1,k_3\to 0) &=& -\frac{3\epsilon}{2}P_{\zeta}(k_1)P_{\zeta}(k_3),\\
B_{\zeta}^{\rm redef}(k_1,k_2\to k_1,k_3\to 0) &=& 2(\epsilon+\delta)P_{\zeta}(k_1)P_{\zeta}(k_3).
\ee
Other terms are sub-dominant in this limit, and thus we get 
\be
B_{\zeta}(k_1,k_2\to k_1,k_3\to 0) &=& (1-n_s)P_{\zeta}(k_1)P_{\zeta}(k_3),
\ee
where we have used Eq.~(\ref{tilt}).
In other words, Eq.~(\ref{consistency_fnl}) is satisfied.

It is curious to note that only the gauge transformation part contributes to the bispectrum in the squeezed limit. 
If $\zeta$ is conserved outside the horizon, then the $\delta N$ formalism gives nothing but the gauge transformation from flat slicing to uniform-field slicing (see \cite{Naruko:2011zk, Gao:2011mz} for conditions that the nonlinear $\zeta$ is conserved in a general class of single scalar field theories). As a result, it immediately gives a complete expression of the local $f_{NL}$ in GEF.

We will now consider a specific model of GEF inflation in which the inflaton is a pseudo-Nambu-Goldstone boson and where the inflaton potential is generated quantum mechanically. In this case, as we shall see, the inflationary scenario is protected under quantum (gravity) corrections.

\section{UV-protected natural inflation}

In natural inflation \cite{Freese:1990rb}, the field $\phi$ is a pseudo-Nambu-Goldstone scalar field with decay constant $f$ and periodicity $2\pi$. Inspired by this idea, we will consider the following tree-level Lagrangian for a single pseudo-scalar field $\phi$
\be\label{axion}
S=\int d^4x\sqrt{-g}\Bigg[\frac{M_p^2}{2} R-\frac{1}{2}\Delta^{\alpha\beta}\partial_\alpha
\phi\partial_\beta \phi-me^{i\frac{\phi}{f}}\bar\psi(1+\gamma_5)\psi -\bar\psi{\not} {\cal D} \psi-\frac{1}{2}\mbox{Tr}F_{\alpha\beta}F^{\alpha\beta}\Bigg] ,
\ee
where $\psi$ is a fermion charged under the (non-abelian) gauge field with field strength $F_{\alpha\beta}$, ${\not} {\cal D} =\gamma^\alpha{\cal D} _\alpha$ is the gauge invariant derivative and $m\sim f$ is the fermion mass scale after spontaneous symmetry breaking.

The action (\ref{axion}) is invariant under the chiral (global) symmetry $\psi\rightarrow e^{i\gamma_5\alpha/2}\psi$, where $\alpha$ is a constant. This symmetry is related to the invariance under shift symmetry of $\phi$, i.e. $\phi\rightarrow\phi-\alpha\ f$.

The chiral symmetry of the system is, however, broken at one loop level \cite{Peskin:1995ev} giving the effective interaction $\frac{\phi}{f} F\cdot \tilde F$, where $\tilde F^{\mu\nu}=(1/\sqrt{-g})\epsilon^{\alpha\beta\mu\nu}F_{\alpha\beta}$ and $\epsilon^{\alpha\beta\mu\nu}$ is the Levi-Civita antisymmetric symbol. Instanton effects related to the gauge theory of field strength $F$ introduce a potential $K(F\cdot\tilde F)$ \cite{Dvali:2005an}. In the zero momentum limit, we can integrate out the combination $F\cdot\tilde F$ and obtain a periodic potential for the field $\phi$ (note that this is independent upon the canonical normalization of $\phi$) which has a stable minimum at $\phi=0$ \cite{Vafa:1984xg}. 

We will now discuss two different regimes of the UV-protected natural inflation.

\subsection{Small field branch}

If we expand the potential around its own maximum, we get
 \be\label{pot}
 V(\phi)\simeq\Lambda^4\left(2-\frac{\phi^2}{2f^2}\right)\ ,
 \ee
 where $\Lambda$ is the strong coupling scale of the gauge theory of field strength $F$ \cite{Gross:1980br}. The approximation (\ref{pot}) is valid as long as $\phi\ll f$ and it is precisely in this regime that the Universe can naturally inflate.
 
With the help of equations (\ref{sr}) we find the following independent conditions extracted from (\ref{srpv}):
\be\label{slowroll23}
\epsilon\simeq\frac{M^2}{24H^2}\frac{\phi^2}{f^2}\frac{M_p^2}{f^2}\ll 1\ ,\quad  |\eta|\simeq\frac{M^2}{6H^2}\frac{M_p^2}{f^2}\ll1\ ,\quad  \frac{M^2}{H^2}\ll 1\ .
\ee
Note that both $\eta$ and $\epsilon$ are suppressed by the additional gravitational friction term $H^2/M^2\gg 1$ that is not present in the original natural inflation \cite{Freese:1990rb}. This enhanced gravitational friction is the key physical mechanism allowing $f\ll M_p$. 

We firstly impose the weak coupling constraint of the gauge interaction with the inflaton, $f\gg M$ (i.e., $\tilde{f}\sim fH/M \gg H$). The quantum gravity constraint such that the curvature should be smaller than the Planck scale\footnote{Or smaller than $\Lambda\simeq(M_p H^2)^{1/3}$, in case of which symmetry is broken by an extra field $\rho$.} is easily satisfied for $\Lambda \ll M_p$ (i.e., $R\sim H^2 \sim \Lambda^4/M_p^2 \ll M_p^2$). The friction constraint $H^2\gg M^2$ is satisfied for $\Lambda^4\gg M^2 M_p^2$, which implies $M \ll M_p$ as it should. Finally, we would like to impose $f\ll M_p$ to avoid trans-Planckian masses. 

For specific models of symmetry breaking, one should also impose the mass of the mode restoring the symmetry (let us call it $\rho$) to be larger than $H$, in order to not excite this mode. For example, in axionic models, this constrains the ratio $f/H$  to be large (during inflation). It is easy to convince ourselves that this constraint is very weak as the larger $f\ (\ll M_p)$ is, the better the slow roll conditions are satisfied. We will therefore disregard this constraint in the following.

Collecting all conditions and constraints, the natural inflationary set-up is UV-protected if the following hierarchies of scales are satisfied:
\be\label{conditions}
M \ll M\frac{M_p^2}{\Lambda^2} \ll f \ll M_p,
\ee
where specifically, the upper bound on $f$ is to protect the flatness of the potential from quantum gravity UV corrections, while the lower bound is to protect it from gauge interaction UV corrections (see paragraphs below Eq.~(9) of \cite{Germani:2010hd}).

The number of efolds in this model is given by
\be\label{efold_uv}
N_e = \int^{\phi_f}_{\phi_i} \frac{H}{\dot\phi}d\phi
\simeq \frac{4\Lambda^4f^2}{M_p^4M^2}\ln{\frac{\phi_f}{\phi_i}},
\ee
where $\phi_f$ is the field value at the end of inflation while $\phi_i$ is the value at the observational scales leave the horizon.
From Eq.~({\ref{tilt}}),
\be\label{tilt_uv}
n_s - 1 \simeq \frac{M^2}{H^2}M_p^2 \Bigg[ -\frac{1}{3}\frac{\phi_i^2}{f^4}-\frac{1}{3f^2} \Bigg]
\simeq -\frac13\frac{M^2}{H^2}\frac{M_p^2}{f^2},
\ee
where we have used $\phi_i \ll f$.
Thus, the UV-protected inflation predicts the red tilted spectrum.
Combining Eqs.~(\ref{efold_uv}) and (\ref{tilt_uv}), one finds a relation
\be\label{relation_phi}
\phi_f = \phi_i e^{N_e(1-n_s)/2}.
\ee
If $N_e=50$ and $1-n_s =0.04$, we have $\phi_f = e\phi_i \ll f$, which justifies the approximation of the potential around a local maximum.

Now, we shall constrain the model by observations. If we match the tilt with the WMAP data \cite{Komatsu:2010fb},
\be\label{ns_wmap}
n_s - 1 = -0.04,
\ee
we get a relation
\be\label{relation_tilt}
\frac{M}{H} = \frac{\sqrt{3}}{5}\frac{f}{M_p}.
\ee
Combining with the constraint of the amplitude~(\ref{relation_ps}), we get another relation
\be
\frac{\Lambda^2}{M_p^2} = \frac{\pi\sqrt6}{10^5\sqrt5}\frac{\phi_i}{f}.
\ee

These relations are consistent with the hierarchy of scales to avoid strong coupling in the model (\ref{conditions}).

Finally, we  discuss about detectability of primordial gravitational waves.

The Lyth bound~(\ref{lyth}) implies
\be
\left( \frac{r}{0.1}\right)^{1/2} \lesssim \frac{(e^{N_e/50}-1)}{4\sqrt3}\frac{\phi_i }{f}\frac{50}{N_e},
\ee
where we have used relations~(\ref{relation_phi}), (\ref{ns_wmap}) and (\ref{relation_tilt}).
Since $\phi_i \ll f$ near a local maximum of the potential, the tensor to scalar ratio, $r$, is negligibly small in this region.

\subsubsection{Infra-red completion?}
Far away after inflation, when the system relaxes to the Minkowski background, we have $f > \Lambda_{flat}=(M^2M_p)^{1/3}$ to be consistent with the observational constraints that we have obtained above. In this case, the perturbative unitarity restoring field $\rho$, cannot be frozen anymore at the scale $f$, unless strongly coupled.\footnote{We thank Fedor Bezrukov for pointing this out.} Therefore, in the Minkowski background, we need to integrate in the $\rho$ field at least. This interesting ``infra-red'' completion of our theory is left for future work.

\subsection{Large field branch}\label{sub:large-uv}

As discussed before, the small field branch of the UV-protected inflation, which is the original model presented in \cite{Germani:2010hd} does not produce any detectable gravitational wave signal. It is then interesting to study the large field branch of the model (\ref{axion}). Indeed, if we expand around the minimum the scalar potential, we have a chaotic like form
\be\label{712}
V(\phi)\simeq \frac{1}{2}m^2\phi^2\ ,\quad m\equiv \frac{\Lambda^2}{f},
\ee
where the approximation is valid for $\phi\ll \pi f$. In this case, one finds
\be\label{sr-uv-large}
\epsilon=\eta=4\frac{M^2 M_p^4}{\phi^4 m^2}\ .
\ee
Imposing the slow roll ($\epsilon\ll 1$), periodicity ($\phi\ll f$) conditions and $f \ll M_p$, we have
\be\label{conlarge0}
\frac{M_p}{\Lambda}\sqrt{M f}\ll \phi \ll f \ll M_p\ .
\ee
Avoidance of quantum gravity regime ($\Lambda \ll M_p$) combined with (\ref{conlarge0}) automatically implies weak coupling of the gauge field-inflaton system during inflation ($f\gg M$) and $H,\ M \ll M_p$.
The above constraints differ from (\ref{conditions}) just by the lower bound for the inflaton field value.

With the definition of the number of e-folds $N_e=\int dt H$, we also find
\be
N_e=\frac{3}{2(1-n_s)}\ ,
\ee
where $n_s$ is defined in Eq.~(\ref{tilt}). 

In order to obtain exactly 50 e-folds, we get $1-n_s=0.03$ as a prediction; this is within the bound of WMAP \cite{Komatsu:2010fb}. This value of the spectral index implies
\be\label{phi-ns-uv-large}
\phi_i^4=8\times 10^2\frac{M^2 M_p^4}{m^2}\ ,
\ee
whereas the normalization of the amplitude of fluctuation (\ref{relation_ps}) implies
\be
\phi_i^6=384\pi^2\times 10^{-9}\frac{M^2 M_p^8}{m^4}\ .
\ee
Note that the above values easily satisfy the constraints~(\ref{conlarge0}) and $\Lambda \ll M_p$.

As this model is effectively a large field model ($\tilde\phi\sim (H/M)\phi\gg M_p$), there are detectable signals of gravitational waves. In fact, here we obtain
\be
r\simeq 0.08\ ,
\ee
where we have used Eqs.~(\ref{ratio}), (\ref{sr-uv-large}) and (\ref{phi-ns-uv-large}).
One might be still worried about the fact that in this regime the scalar field covers a trans-Planckian range from the beginning to the end of inflation. However, as quantum gravity correction must respect the discrete symmetry of the system, they can only slightly modify $\Lambda$, as discussed in \cite{Germani:2010hd}.

\subsubsection{Saving $\lambda\phi^4$ model}\label{subsub:phi4}

We now move slightly away from the main focus of this paper by considering non-UV protected scenarios and discuss the peculiar case of the $\lambda\phi^4$ model. Indeed, although this model, in its canonical realization, has been excluded by observations \cite{Komatsu:2010fb}, its GEF version turns out to be compatible with the observational constraints. The New Higgs inflation of \cite{Germani:2010gm} is a physically motivated example using this potential with the GEF mechanism.

The $\lambda\phi^4$ model predicts a red spectrum \cite{Germani:2010ux}:\footnote{Although the lapse of \cite{Germani:2010ux} [Eq.~(3.6)] is missing a factor of $\Gamma$ [see Eq.~(\ref{lapse}) of this paper],  $n_s-1$ agrees with Eq.~(3.18) of \cite{Germani:2010ux}.}
\be\label{tilt_higgs}
V=\frac{\lambda}{4}\phi^4,\quad
n_s - 1  \simeq -\frac{40}{3}\frac{M^2}{H^2}\frac{M_p^2}{\phi_i^2}\simeq -5\epsilon
\ee
 and
\be
N_e = \frac{5}{3(1-n_s)}.
\ee

For $n_s-1= -0.03$, one obtains 
\be\label{values_phi4}
\epsilon &\simeq& 0.006,\ 
N_e \simeq 56,\
r \simeq 0.1
\ee
and
\be
\frac{\phi_i}{M_p}&\simeq& 0.018\left(\frac{0.1}{\lambda}\right)^{1/4},\ 
\frac{H}{M_p}\simeq 2.4\times 10^{-6},\ 
\frac{M}{M_p}\simeq 2.7\times 10^{-8}\left(\frac{0.1}{\lambda}\right)^{1/4},
\ee
where Eqs.~(\ref{sr}), (\ref{relation_ps}), (\ref{ratio}) and (\ref{tilt_higgs}) have been used.
Values~(\ref{values_phi4}) are compatible with the observations \cite{Komatsu:2010fb}.

\subsection{Non-Gaussianity from gauge interaction with the pseudo-scalar inflaton}\label{sec:barnaby-peloso}

Non-Gaussianity can be generated by the inverse decays of gauge fields if the inflaton is identified as a pseudo-scalar (Barnaby-Peloso mechanism \cite{Barnaby:2010vf, Barnaby:2011vw}).

In principle, the inflaton may also couple to any gauge boson with field strength $F_i$, with decay constant $f_i$.

In this case, the generation of gauge field fluctuations is governed by a set of parameters \cite{Anber:2009ua, Barnaby:2010vf}
\be
\xi_i \equiv \frac{\dot\phi}{2f_iH}=\xi \frac{f}{f_i}\ ~~{\rm where}~~\ \xi\equiv \frac{\dot\phi}{2fH}\ .
\ee
If $\xi_i \gtrsim {\cal O}(1)$, the positive helicity mode of the gauge field is amplified exponentially \cite{Anber:2009ua}, and generates non-Gaussian curvature perturbations at second order.

Barnaby and Peloso have found \cite{Barnaby:2010vf} 
\be
f_{NL}^{\rm equil} \simeq 4.4\times 10^{10}{\cal P}_{\zeta}^3\frac{e^{6\pi\xi_i}}{\xi_i^9}.
\ee
At sufficiently large $\xi_i \gtrsim {\cal O}(1)$, $f_{NL}^{\rm equil}\simeq8400$, which excludes axion-like inflation models by the observations \cite{Komatsu:2010fb}.

In the small field branch of the UV-protected inflation, 
\be
\xi \equiv \frac{\dot\phi}{2fH} \simeq \sqrt{\frac{\epsilon}{6}}\frac{M}{H}\frac{M_p}{f} = \frac{\sqrt{\epsilon}}{5\sqrt2}\simeq 2\times 10^{-2} \ll 1,
\ee
where we have used the relation~(\ref{relation_tilt}).
The factor $M/H$ makes $\xi$ less than ${\cal O}(1)$, and thus there is no observational signature of non-Gaussianity produced by the Barnaby-Peloso mechanism from the gauge interaction producing the inflaton potential. Once again we can explain this suppression very easily. During the high friction limit indeed, the effective decaying constant for the canonically normalized scalar field becomes $\tilde f\sim f H/M \gg M_p$ so that $\xi\sim M_p/\tilde f \ll 1$.\footnote{The scale $\tilde{f}$ is the effective gauge coupling constant with the inflaton. It would set the strong coupling scale but is not related to any physical mass scale; thus it can be larger than the Planck scale.} 
The same conclusion can be reached for the large field branch. 

The Barnaby-Peloso mechanism implies also new constraints for other gauge interactions with the inflaton. 
These extra fields may be Abelian and/or non-Abelian (with trivial and/or non-trivial vacua) gauge fields.
As discussed before, we indeed have\footnote{One can also find a similar relation for large field regime, however here there is considerable freedom in choosing the value of $\xi$.}
\be
\xi_i\simeq2\times 10^{-2}\ \frac{f}{f_i}\ .
\ee
Current bounds on non-Gaussianities require $\xi_i\lesssim 2.6$ \cite{Barnaby:2010vf}, so that for a detectable signal we need
\be\label{large-fnl-f}
f_i\sim 10^{-2}\ f\ ,
\ee
where the smaller values of $f_i \lesssim 10^{-2} f$ are tightly constrained by the observations.
It is also necessary to avoid strong couplings of the inflaton with additional gauge bosons during inflation; thus, canonically normalizing the inflaton in the high friction limit, we should have
\be\label{before}
\frac{H}{M} f_i \gg H\ ,{\rm namely},\ \ f_i\gg M\ .
\ee
In addition, for gauge fields with non-trivial vacuum, we also need to bound the effective mass of $\phi$ due to these extra interactions to be 
\be
\frac{\Lambda^2_i}{f_i}\ll \frac{\Lambda^2}{f}\ ,
\ee
where $\Lambda_i$ is the strong coupling scale of any gauge field with field strength $F_i$ that produces the instanton effect.

\section{Conclusions}
 
 The GEF mechanism is a very powerful way to increase friction of a scalar field rolling down its own potential without introducing any new degree of freedom. In this way, virtually any scalar field potential is able to produce successful inflation without violating the perturbative unitarity bound of the theory. 
With the GEF mechanism one can, for example, revive the $\lambda\phi^4$ inflation as an observationally viable model, as we discussed in the subsection~\ref{subsub:phi4}.

This mechanism, in order to work, needs a nonminimal coupling of the Einstein tensor to the kinetic term of the inflaton.
Although one may be tempted to say that non-Gaussianities will be boosted by this nonlinear interaction, for example in the squeezed limit, they actually work  the opposite whenever the GEF is efficient. In the high friction limit indeed, non-Gaussian fluctuations of the scalar field vanish at order $\epsilon$. 
In this respect, the non-Gaussian contribution is completely dominated by the nonlinear gauge transformation from the spatially flat to uniform-field gauge. 
In the uniform-field gauge ($\delta\phi=0$) indeed, everything behaves as in canonical GR. 
This is due to the fact that, in this gauge, the nonlinearity, and the new scale $M$, are completely absorbed into redefinitions of the slow roll parameters.
We have explicitly showed a consistency relation between the bispectrum in the squeezed limit and the spectral tilt in the subsection~\ref{sub:local}.

These generic features are also used to constrain the so-called UV-protected inflation \cite{Germani:2010hd}; this is our main focus. The UV-protected inflation is realized with a pseudo-scalar with a potential generated by quantum one-loop breaking of a global symmetry into a discrete one. Thanks to the GEF, inflation is then achieved with sub-Planckian parameter scales for the inflaton Lagrangian, making the UV-protected inflation of \cite{Germani:2010hd}, insensible from UV quantum (gravity) corrections.

We have showed that this model predicts a red tilted spectrum of primordial curvature perturbations and possible gravitational wave detections.

In addition, we have showed that extra couplings of the inflaton to other gauge fields weakly participating to the inflaton potential, may produce detectable non-Gaussian signals via the mechanism of \cite{Barnaby:2010vf}.

Concluding, we would like to briefly mention about reheating after inflation and postpone this important analysis for future work. There are two possibilities: the first one is that the GEF interaction becomes subdominant after inflation due to the rapid decrease of the Hubble constant, in this case reheating works similarly as minimally coupled inflationary scenarios. The second possibility and perhaps the more interesting, makes actually use of the GEF mechanism. In this case, after diagonalizing the scalar and graviton degrees of freedom coupled via the GEF interaction, one obtains, as a result, an effective inflaton coupling to the standard model particles. This is the gravitational inflaton decay mechanism studied for example in \cite{Vilenkin:1985md,Kalara:1990ar,Watanabe:2006ku,Watanabe:2010vy} for conformal type nonminimal couplings.

\acknowledgments
The authors wish to thank Alex Kehagias and Gerasimos Rigopoulos for interesting discussions and for an early participation to this work. They thank Neil Barnaby, Jonathan Ganc, Eiichiro Komatsu, Andrei Linde, Misao Sasaki and Lorenzo Sorbo for valuable comments on an earlier version of the draft. They would also like to thank Fedor Bezrukov for discussions on the physical validity of the model and Nico Wintergerst for helps in using XTensor. They acknowledge the use of Mathematica (XTensor) and Maple. CG is supported by Humboldt Foundation. YW is supported by the TRR 33 "The Dark Universe". This work is partially supported by the PEVE-NTUA-2009 program. 

\bibliographystyle{jhep}
\bibliography{watanabe}
\end{document}